\documentclass[12pt,epsf,axodraw]{article}

\input epsf.sty

\newcommand{\bea}{\begin{eqnarray}}
\newcommand{\eea}{\end{eqnarray}}
\newcommand{\be}{\begin{equation}}
\newcommand{\ee}{\begin{equation}}

\begin{document}

\thispagestyle{empty}

\begin{flushright}
{\tt  VEC/PHYSICS/P/2/2004-2005}
\end{flushright}

\begin{center}
{\LARGE \bf Upper bound on the mass scale of superpartners
in minimal $N=2$ supersymmetry \\}
\vspace{2cm}
{\bf Biswajoy Brahmachari}
\\
\vskip 1cm
{\sl Department of Physics, Vidyasagar Evening College,\\
39, Sankar Ghosh Lane, Kolkata 700006, India.\\}
\end{center}
\vskip 1cm
\begin{abstract}

If $N=2$ supersymmetry breaks to $N=1$ supersymmetry at an
intermediate scale $m_2$ and then, later on, $N=1$ supersymmetry 
breaks and produces standard model at a scale $m_{susy}$ such that 
$m_2 > m_{susy}$, renormalization group evolution of three gauge
couplings are altered above the scale $m_2$, changing the unification 
scale and the unified coupling. We show that when we enforce this
general condition $m_2 > m_{susy}$ on the solutions of the
renormalization group equations, the condition is translated into an 
upper bound on the scale $m_{susy}$. Using presently favored values of 
$\alpha_1(m_z),\alpha_2(m_z),\alpha_3(m_z)$, we get $m_{susy} < 4.5 
\times 10^9$ GeVs for the central value of $\alpha_3(m_Z)$. 
When low energy threshold effect is present, this bound gets smeared
yet remains generally stable in the $10^9-10^{10}$ GeV range. We also
show that if we demand string 
unification instead of 
having an unified gauge theory, this constraint can be changed by 
exotic hypercharge normalizations.

\end{abstract}

\vskip .75in

\begin{flushleft}
{\tt email: biswajoy.brahmachari@cern.ch}
\end{flushleft}

\newpage

In minimal supersymmetric standard model (MSSM) there exists a fundamental 
scale $m_{susy}$. It is a common mass scale of the superpartners of
Standard Model(SM) fields. We usually assume that it is less than a TeV or so
because it helpful for solving the gauge hierarchy problem. In principle
$m_{susy}$ can be as high as the Planck scale or the String scale. In
this paper we will seek an upper bound on the scale $m_{susy}$ by
demanding that gauge couplings should unify at some scale $M_X$, and
there exists a natural hierarchy of four mass scales of the form 
$M_X > m_2 > m_{susy} > m_z$. Here $m_2$ is the scale where $N=2$ 
supersymmetry breaks, all mirror particles of MSSM become heavy and 
they decouple from Renormalization Group(RG) running. As in usual 
notation $m_{susy}$ is the scale where super partners of Standard 
Model becomes heavy and so they decouple from RG running. Therefore 
we see that $m_{susy}$ is the scale where $N=1$ supersymmetry breaking 
is expected to be directly felt by experiments\cite{ref0-exp1,
ref0-exp2,ref0-exp3,ref0-exp4,ref0-exp5,ref0-exp6,ref0-exp7},
consequently it is quite important to search for any theoretical upper 
bound that may exist on the mass scale $m_{susy}$. 

Supersymmetric extensions of the standard model are interesting
from two points of view. (i) Gauge coupling unification is precise
at the scale of approximately $2 \times 10^{16}$
GeV\cite{ref1-amaldi1,ref1-amaldi2,ref1-amaldi3,ref1-amaldi4}. (ii) 
Divergences in the scalar sector are 
canceled by loops involving superpartners of standard particles which 
helps to solve hierarchy problem partially\cite{ref2-kaul1,ref2-kaul2,
ref2-kaul3}. $N=2$
supersymmetric extensions of the standard model are relatively less
studied\footnote{In extra-dimensional models N=2 supersymmetry occurs
frequently. See for example \cite{ref11-n2ex1,ref11-n2ex2,ref11-n2ex3,
ref11-n2ex4,ref11-n2ex5}} even-though they are 
much more restrictive than the $N=1$ framework. Particularly after the 
breakdown of $N=2$ supersymmetry, vanishing of supertrace $Str(M^2)$ 
condition forces all field dependent quartic divergences to be 
zero\cite{ref3-str1,ref3-str2,ref3-str3,ref3-str4}, which is a
desirable ingredient for solving the 
hierarchy problem in a more comprehensive manner. There are known
mechanisms by which $N=2$
supersymmetry can be spontaneously broken down to $N=1$ in local
quantum field theories. It is therefore relevant to consider a
symmetry breaking chain in which $N=2$ supersymmetry is spontaneously 
broken at an intermediate scale below the unification scale. This 
possibility is studied by Antoniadis, Ellis and Leontaris (AEL)
\cite{ref4-ael}. In their analysis AEL assumed $m_Z \equiv m_{susy}$,
or in other words minimal supersymmetric standard model is effective
very near the electroweak scale. Consequently they used $N=1$ beta 
functions from the electroweak scale and the intermediate scale and 
$N=2$ beta functions from the intermediate scale to the unification 
scale. In this present analysis we separate $m_Z$ from $m_{susy}$ and 
make $m_{susy}$ a free parameter and then we run couplings up to the 
scale $m_{susy}$ using non-supersymmetric beta functions, from
$m_{susy}$ to $m_2$ we use $N=1$ supersummetric beta functions as was 
done by AEL and from $m_2$ to the unification scale $M_X$ we use $N=2$ 
supersymmetric beta functions exactly as AEL performed. We aim to
obtain constraints on the scale $m_{susy}$ enforcing two conditions 
(a) gauge coupling unification should take place (b) The condition 
$m_2 > m_{susy}$ has to be satisfied. We will see that in this way we 
can obtain an interesting upper bound on $m_{susy}$. This the result 
that we are reporting in this paper.

We know that if $N=4$ supersymmetry is present, beta functions vanish 
at all orders\cite{ref5-n41,ref5-n42,ref5-n43,ref5-n44,ref5-n45}, but if $N=2$ supersymmetry is present, they
vanish beyond one-loop 
order\cite{ref6-n21,ref6-n22,ref6-n23,ref6-n24,ref6-n25}. For $N=1$ 
supersymmetry, 
however, using one loop beta functions is an approximation. This 
approximation is justified in the context of the present analysis. Had 
we done a precision test of whether gauge couplings are at all
unifying or not in a restrictive scenario like unification in MSSM it 
would have been necessary to use higher loop beta functions. However,
our objective is not to do a precision test of gauge coupling unification. We 
will give an upper bound on the scale $m_{susy}$ is a theory with 
two intermediate scales namely $m_{susy}$ and $m_2$; one does not gain
appreciably in precision by using higher loop beta functions in a
theory with many unknown intermediate scales. Using two loop
beta function below $m_2$ will give a slight shift in the values 
of $m_{susy}$ and $m_2$. Because we have to cross two thresholds $m_2$ and
$m_{susy}$, and we are neglecting unknown threshold effects at those
intermediate scales, it is reasonable to neglect two-loop corrections to gauge
coupling evolution which is comparable to these threshold corrections.

The three gauge couplings evolve via the Renormalization Group
Equations (RGE). The solutions of RGE in the energy range $m_Z
\longrightarrow m_{susy} \longrightarrow m_2 \longrightarrow M_X$ are,
\begin{eqnarray}
\alpha^{-1}_1(m_Z) &=& \alpha_X^{-1}+ 
{\beta^{N=2}_1 \over \kappa}~\ln{M_X \over m_2}+
{\beta^{N=1}_1 \over \kappa}~\ln{m_2 \over m_{susy}}+
{\beta_1 \over \kappa}~\ln{m_{susy} \over m_Z}, \label{sol1}\\
\alpha^{-1}_2(m_Z) &=& \alpha_X^{-1}+ 
\beta^{N=2}_2~\ln{M_X \over m_2}+
\beta^{N=1}_2~\ln{m_2 \over m_{susy}}+
\beta_2~\ln{m_{susy} \over m_Z}, \label{sol2}\\
\alpha^{-1}_3(m_Z) &=& \alpha_X^{-1}+ 
\beta^{N=2}_3~\ln{M_X \over m_2}+
\beta^{N=1}_3~\ln{m_2 \over m_{susy}}+
\beta_3~\ln{m_{susy} \over m_Z}\label{sol3}. 
\end{eqnarray} 
Here $\alpha_X^{-1}$ is the inverse of unified coupling, $\kappa$ is 
the $U(1)$ normalization factor, which is usually taken as ${5 \over
3}$ valid for the $SU(5)$ case, but could be different as well in 
string models, and the beta coefficients without U(1) normalizations 
are listed in Table \ref{table1}.

\begin{table}
\begin{tabular}{||c|c|c||c|c|c||c|c|c||} 
\hline
\multicolumn{3}{|c|}{N=2 SUSY}
&
\multicolumn{3}{|c|}{N=1 SUSY} 
&
\multicolumn{3}{|c|}{NON-SUSY}
\\
\cline{1-3}
\cline{4-6}
\cline{7-9}
$2 \pi \beta^{N=2}_3$ & $2 \pi \beta^{N=2}_2 $& $2 \pi \beta^{N=2}_1$ & 
$2 \pi \beta^{N=1}_3$ & $2 \pi \beta^{N=1}_2 $& $2 \pi \beta^{N=1}_1$ & 
$2 \pi \beta_3$ & $2 \pi \beta_2 $& $2 \pi \beta_1$ 
\\
\hline
6 & 10 & 22 & -3 & 1 & 11 & -7 & -19/6 & 41/6 \\
\hline
\end{tabular}
\caption{ This table gives beta coefficients without the U(1)
normalization factor. When we divide columns 3,6,9 by 5/3 we 
get, 66/5, 33/5 and 41/10 which are well-known values in SU(5) case.}
\label{table1}
\end{table}

Let us use canonical normalization $\kappa={5 \over 3}$, and the
values of three gauge couplings at the scale $m_Z$ to be
$\alpha_1(m_Z)=0.01688$, $\alpha_2(m_Z)=0.03322$, and $\alpha_3(m_Z)=0.118$.
We solve three simultaneous equations Eqn. \ref{sol1}-\ref{sol3} for
three unknowns. The unknowns are
$\alpha^{-1}_X, \ln{m_2 \over m_{susy}},  \ln{M_X \over m_2}$.
After solving we find that,  
\begin{eqnarray}
\alpha^{-1}_X &=& 22.42 - 0.48 ~~\ln{m_{susy} \over
 m_Z},\label{bound11}
 \\
\ln{m_2 \over m_{susy}} &=& 31.00 - 1.75 ~~\ln{m_{susy} \over m_Z},
\label{bound12}\\
\ln{M_X \over m_2} &=& 2.98 + 0.79~~\ln{m_{susy} \over m_Z}
\label{bound13}.
\end{eqnarray}

\begin{figure}
\begin{center}
\epsfysize=13cm \epsfxsize=13cm \epsffile{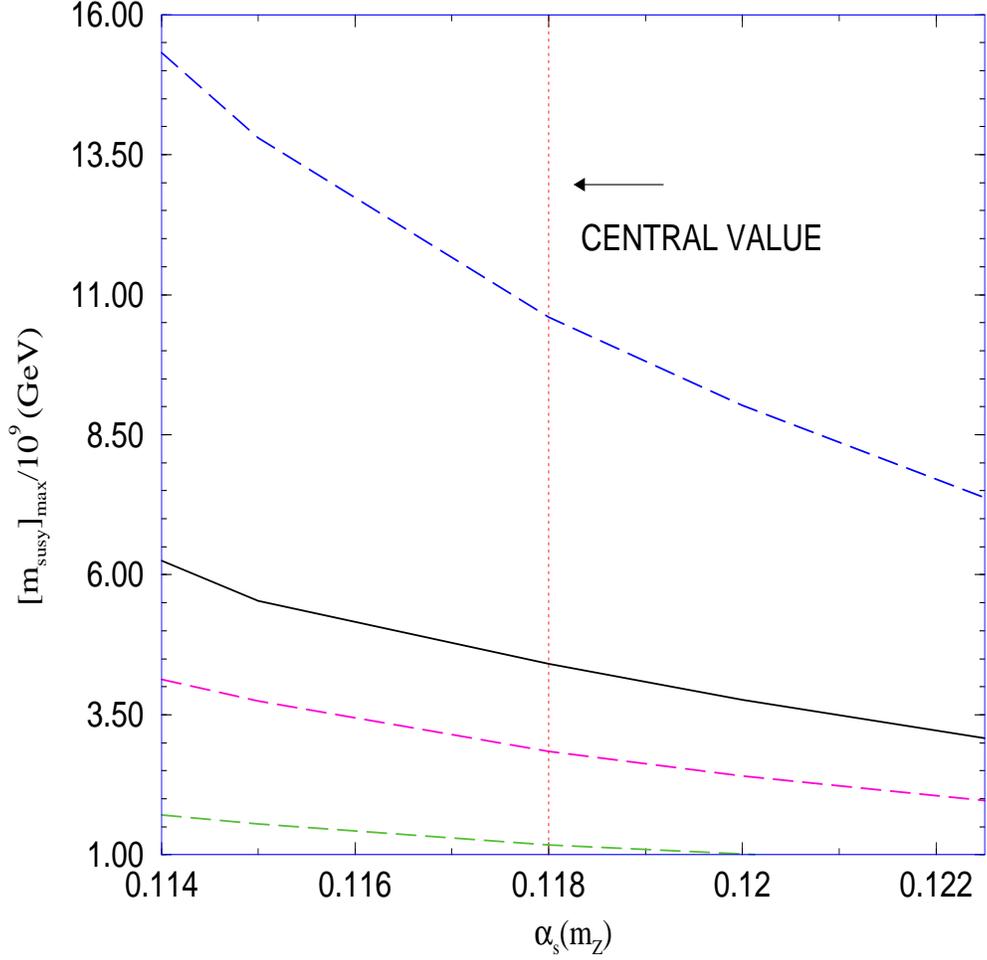} \caption{
Upper bound on $m_{susy}$ for canonical U(1) normalization.
Dashed magenta line shows effects of threshold corrections when wino
and gluino thresholds are one order of magnitude higher than
$m_{susy}$. Blue dashed line is the case where wino mass is the
same as $m_{susy}$ but gluino mass is higher than $m_{susy}$ by one
order of magnitude. Green dashed line is the case when gluino mass
is the same as $m_{susy}$ but wino mass is larger by one order of
magnitude.
} \label{fig1}
\end{center}
\end{figure}

Because $m_2 > m_{susy}$, 
$\ln{m_2 \over m_{susy}}$ has to be non-negative.  Therefore, from
Eqn. \ref{bound12} we see that  
$\ln{m_{susy} \over m_Z}$ 
has an absolute
upper bound at  $31.00/1.75=17.71$. Using the value $m_Z=91.2$ GeV
we find that
\begin{equation}
m_{susy} < 4.48 \times 10^9~~~{\rm GeV}\label{result}.
\end{equation}
This upper bound depends on the value of $\alpha_3(m_Z)$. 
We have plotted this upper bound in solid black line Fig. \ref{fig1} 
for values of
$\alpha_3(m_Z)$ in the range $0.11-0.13$ and for the canonical U(1)
normalization of $\kappa=5/3$. For the central value of
$\alpha_3(m_Z)=0.118$, and $m_{susy}$ at its upper limit, we get
$\alpha^{-1}_X= 12.10, M_X=1.02 \times 10^{17}$ GeV. For 
$m_{susy} \approx m_Z$, we get, for
$\alpha_3(m_Z)=0.118$, three solved quantities to be, 
$\alpha^{-1}_X=20.42, m_2= 2.6 \times 10^{15}$ GeV and $M_X= 5.1 \times
10^{16}$ GeV. Consequently, we see that, we reproduce
results obtained by AEL in the special case of $m_Z \approx m_{susy}$  as
expected.  

Now let us discuss low energy threshold effects and how it may
influence our results. To see this let us recast the RGE and
include two more thresholds, namely the, $M_{\tilde{g}}$ and
$M_{\tilde{w}}$. These two are the most important supersymmetric
thresholds because the gauge contribution to beta function coefficients 
dominates over fermion contribution and Higgs contribution.
Evolution of $\alpha_1$ remains unaffected at one-loop because
it does not have to cross any new threshold. 
When threshold effect is present $M_{\tilde{g}}$ 
and $M_{\tilde{w}}$ are different from the common mass scale
$m_{susy}$. Unfortunately we do not know how different they 
actually are. We can use a few representative cases only. 
The RGE 
now becomes,
\begin{eqnarray}
\alpha^{-1}_2(m_Z) &=& \alpha_X^{-1}+ 
\beta^{N=2}_2~\ln{M_X \over m_2}+
\beta^{N=1}_2~\ln{m_2 \over m_{\tilde{w}}}+
(\beta^{N=1}_2-\Delta_{w})~\ln{m_{\tilde{w}} \over m_{susy}} 
\nonumber\\
&& +\beta_2 ~\ln{m_{susy} \over m_z}
, \label{sol2p}\\
\alpha^{-1}_3(m_Z) &=& \alpha_X^{-1}+ 
\beta^{N=2}_3~\ln{M_X \over m_2}+
\beta^{N=1}_3~\ln{m_2 \over m_{\tilde{g}}}+
(\beta^{N=1}_3-\Delta_{g})~\ln{m_{\tilde{g}} \over m_{susy}} 
\nonumber\\ 
&& +\beta_3 \ln{m_{susy} \over m_Z}
\label{sol3p}. 
\end{eqnarray} 

We will use $\Delta_g=2$ and $\Delta_w=4/3$. An easy way to see
these numbers is the following. In the non-supersymmetric cases gauge
contribution to $SU(3)$ and $SU(2)$ beta functions are $-11$ and
$-22/3$ respectively. If we add to them the gluino contribution
$\Delta_g=2$ and wino contribution $\Delta_w=4/3$ we get $-9$ and $-6$
which are gauge contributions in the supersymmetric case. 
Using values given in Table \ref{table1}, now we get,
\begin{eqnarray}
2 \pi (\beta^{N=1}_3-\Delta_g) &=& -5 \label{d1}\\
2 \pi (\beta^{N=1}_2-\Delta_w) &=& -{1 \over 3} \label{d2}
\end{eqnarray}
Furthermore let us use two more identities,
\begin{equation}
\ln {m_2 \over m_{susy}}=
\ln {m_2 \over m_{\tilde{w}}} + \ln
{m_{\tilde{w}} \over m_{susy}}
=\ln {m_2 \over m_{\tilde{g}}} + \ln
{m_{\tilde{g}} \over m_{susy}}
\end{equation}
Then the RGE can be rewritten as,
\begin{eqnarray}
\alpha^{-1}_2(m_Z) &=& \alpha_X^{-1}+ 
\beta^{N=2}_2~\ln{M_X \over m_2}+
\beta^{N=1}_2~
(\ln{m_2 \over m_{susy}}-\ln{ m_{\tilde{w}} \over m_{susy}})
\nonumber\\
&& +
(\beta^{N=1}_2-\Delta_{w})~\ln{m_{\tilde{w}} \over m_{susy}} 
+\beta_2 ~\ln{m_{susy} \over m_z}
, \label{sol2pp}\\
\alpha^{-1}_3(m_Z) &=& \alpha_X^{-1}+ 
\beta^{N=2}_3~\ln{M_X \over m_2}+
\beta^{N=1}_3~
(\ln{m_2 \over m_{susy}}-\ln{ m_{\tilde{g}} \over m_{susy}})
\nonumber\\
&& +(\beta^{N=1}_3-\Delta_{g})~\ln{m_{\tilde{g}} \over m_{susy}} 
 +\beta_3 \ln{m_{susy} \over m_Z}
\label{sol3pp}. 
\end{eqnarray} 
Now let us define two threshold parameters ($\sigma_w,\sigma_g$) which
vanish in the limit $m_{susy}=m_{\tilde{w}}=m_{\tilde{g}}$.
\begin{equation}
\ln {m_{\tilde{w}} \over m_{susy}}=\sigma_w,~~
\ln {m_{\tilde{g}} \over m_{susy}}=\sigma_g, \label{d3}
\end{equation}
Magnitude of these parameters are roughly of the order of $\ln_e
10=2.302$ 
when the gluiono
and wino masses differ from $m_{susy}$ by one order of magnitude.
Then we get corrected RGE after threshold corrections using
Eqn. \ref{d1}, \ref{d2}, \ref{d3},
\begin{eqnarray}
\alpha^{-1}_2(m_Z) &=& \alpha_X^{-1}+ 
\beta^{N=2}_2~\ln{M_X \over m_2}+
\beta^{N=1}_2~
(\ln{m_2 \over m_{susy}}-\sigma_w)
\nonumber\\
&& +
\beta_2 ~\ln{m_{susy} \over m_z}-{\sigma_w \over 6 \pi}
, \label{sol2ppp}\\
\alpha^{-1}_3(m_Z) &=& \alpha_X^{-1}+ 
\beta^{N=2}_3~\ln{M_X \over m_2}+
\beta^{N=1}_3~
(\ln{m_2 \over m_{susy}}-\sigma_g)
\nonumber\\
&&  
 +\beta_3 \ln{m_{susy} \over m_Z} -{5 \sigma_g \over 2 \pi}
\label{sol3ppp}. 
\end{eqnarray} 
These equation reduce to Equations \ref{sol2}, \ref{sol3}
in the limit of $\sigma_w \rightarrow 0, \sigma_g \rightarrow 0$. 
So we will solve Equations \ref{sol1},\ref{sol2ppp},\ref{sol3ppp}
to get threshold corrections on our bound. From Fig. \ref{fig1}
we see the results. For $\alpha_s=0.118$ four representative cases can 
be compared.
(i) When all superpartners are degenerate at $m_{susy}$ we get
the bound at $4.5 \times 10^9$ GeV. (ii) When both gluino as well as
wino thresholds are one order of magnitude larger than $m_{susy}$ the
bound becomes $2.85 \times 10^9$ GeV. (iii) When the wino mass is
degenerate with $m_{susy}$ but the gluino mass is one order of
magnitude larger then the bound is $1.06 \times 10^{10}$ GeV. (iv)
When the gluino is degenerate with $m_{susy}$ but the wino mass is one
order of magnitude larger the bound becomes $1.18 \times 10^9$ GeV.
So we see that the bound remains stable in the same ball-park region of 
$10^9-10^{10}$ GeVs when threshold effects are included. The upper
bound is smeared due to threshold correction. This is not surprising
as we know that threshold corrections
have a very similar smearing effect on the mass scales such as the 
unification scale or intermediate scale of all supersymmetric GUTs.

The upper bounds diplayed on Fig \ref{fig1}. should undergo further
small corrections when threshold effects at the N=2 supersymmetry breaking
scale is included. We have not considered heavy threshold effects in the
text because it is beyond the scope of present letter. But a general
observation is that when the spread of n=2  superpartner masses are near
the scale $m_2$ the bound will remain more or less stable near the values
given in Fig \ref{fig1}. The theoretical reason behind it is that the 
mass scales
in the RGE enters only through natural logarithms. Therefore the results
do not get much affected by  small fluctuations in individual masses
near about the scale $m_2$. Note that only new particles beyond
standard model those are included in this analysis have their origin
in $N=1$ and $N=2$ supersymmetry. Therefore their masses must be tied
to  the scales $m_{susy}$ and $m_2$ and fluctuation will remain under control.

Another possiblility is the existance of ad-hoc new thresholds such
as exotic particles between $M_X$ and $m_z$ which are completely unrelated
to N=1 and N=2 supersymmetry. Their masses will therefore be completely
unrelated to $m_{susy}$ or $m_2$. They may change the upperbound
considerably. But here we have not considered exotic new particles which
are not predicted by $N=1$ or $N=2$ supersymmetry.

Furthermore we would like to comment on extra vector-like exotic matter those
may exist anywhere in between $M_X$ and $m_{susy}$. Such vector-like
matter may get masses from the Giudice-Masiero \cite{gm} type mechanism
which is often used to get the mass of the $\mu~H_1~H_2$ term.
Their existence will change the beta functions and alter the present
RGE analysis. Therefore the existence of such extra vectorlike matter
will also change the upper bounds quoted in this paper. Because we
have worked on the minimal version of $N=2$ supersymmetry, we have
not considered exotic vector-like matter either.

Now let us discuss briefly how this upper bound on $m_{susy}$ can be
changed. If we take the canonical value of $\kappa=5/3$ results of 
Fig. \ref{fig1} are obtained. If we notice numerical values of $U(1)_Y$
beta functions we will realize that electric charge is defined as,
\begin{equation}
Q=T^3_L + \sqrt{5 \over 3} Y.
\end{equation}
This is a consequence of the fact that all charges of matter multiplets are
normalized under either $SU(3)$ or $SU(2)_L$ or $U(1)_Y$ in a similar
manner. The underlying assumption being that the generators of $SU(3)$ or
$SU(2)_L$ or $U(1)_Y$ are unified as generators of a bigger unified
gauge group. This is a natural demand if we want gauge coupling
unification as some scale below the mass scale of string theory.
If this is not the case, and string theory breaks directly to $SU(3)_3
\times SU(2)_L \times U(1)_Y$ without passing through an unified
gauge theory just below string scale, the charge relation need not 
have the factor $\sqrt{5 \over 3}$ \cite{ref7-kappa1,ref7-kappa2,
ref7-kappa3,ref7-kappa4,ref7-kappa5} and the general 
string inspired unification condition then reads,
\begin{equation}
K_3 ~\alpha_3=K_2 ~\alpha_2= K_Y~ \alpha_1.
\end{equation}
Let us choose $K_3=K_2=1$ and $K_Y \ne 1$.
Then the upper-bound can be 
changed. If we take a sample value of 
$K_Y= 17/3$, then in this string inspired model, electric charge is
defined as,
\begin{equation}
Q=T^3_L + \sqrt{17 \over 3} Y,
\end{equation}
and, corresponding solution of mass scales reads as,
\begin{eqnarray}
\alpha^{-1}_X &=& 2.04222 - 0.0477465 ~~\ln{m_{susy} \over
 m_Z},\label{boundn1} \\
\ln{m_2 \over m_{susy}} &=& 18.1579 - 1.45 ~~\ln{m_{susy} \over m_Z},
\label{boundn2}\\
\ln{M_X \over m_2} &=& 15.8149 + 0.491667 ~~\ln{m_{susy} \over m_Z}
\label{boundn3}.
\end{eqnarray}
From Eqn. \ref{boundn2} we see that to keep $\ln{m_2 \over m_{susy}}$ 
non-negative we have to have $\ln{m_{susy} \over m_Z}$ less than 
12.5227, which gives $m_{susy} < 2.50 \times 10^{7}$. So the 
bound on $m_{susy}$ given in Eqn. \ref{result} is changed by two 
orders of magnitude. Also note that for $ \ln{m_{susy} \over m_Z}=0$ one
get $\alpha_X=0.4896, m_2=7.01 \times 10^9, M_X=5.17 \times 10^{16}$
thus correctly reproducing numbers quoted by AEL (c.f. Table 3 row 1 of
AEL) for the special case of $m_Z=m_{susy}$.

Let us now discuss two relevent points regarding this letter. (i) This
is a $N=2$ supersymmetric model broken to $N=1$ supersymmetric
model. This may look unfamiliar. However whenever we work on
supersymmetric unification we assume that there is string theory at
some scale above the GUT scale. String theory predicts $N=4$
supersymmetry. So at some stage $N=4$ supersymmetry has to break to
$N=2$ supersymmetry which will then break to $N=1$ supersymmetry.
(ii) The upper bound is very high, which is not attainable at
foreseeable future. This seemingly uninteresting result is relevant
for the following reason. There are plans to probe $N=1$ supersymmetric
particle content in future experiments. Let us assume that we want to search
superpartners at the scale of 40 TeV. We may ask that is that too high
a scale to search for superpartners if $N=2$ supersymmetry breaks to $N=1$?
We have given the answer to that here which states that in the class of
models where the symmetry breaking chain is $N=2 \rightarrow N=1$, if
superpartners exist at (say) 40 TeV, they will not conflict with gauge
coupling unification. Our result also says that if $N=1$ supersymmetry 
is broken at a scale higher than $6.35 \times 10^9$ GeV, we will not 
achieve gauge coupling unification.

In conclusion, we have generalized the analysis of AEL by separating
two scales $m_Z$ and $m_{susy}$. In the paper of AEL the question that
was asked was how much one can lower the scale $m_2$ which is the
scale of $N=2$ breaking and also what other relevant constraints can
be imposed upon $m_2$. In the present analysis we ask the question,
how high the scale $m_{susy}$ can be in that context by unlocking to 
scales $m_Z$ and $m_{susy}$ which were assumed to be equal in the 
analysis of AEL. This analysis is in some sense complementary to the 
analysis of AEL. The scale $m_{susy}$ is very important from the point 
of view of experiments. This is because in experiments we search for
superpartners of standard model particle at around the scale
$m_{susy}$. In principle $m_{susy}$ can be as high as the Planck
scale\cite{ref9-planck1,ref9-planck2,ref9-planck3}, ie, all
superpartners become massive at 
Planck scale and below Planck scale there is non-SUSY standard model. 
In such a case, present day experiments will not be able to trace any 
superpartner. This is the reason why any theoretical or experimental 
upper bound or lower bouund\cite{ref10-lower} on the scale $m_{susy}$, 
that may exist, should be explored.

This work is supported by UGC, New Delhi, under the grant number
F.PSU-075/05-06

\end{document}